# Assessing Long-Distance Atmospheric Transport of Soilborne Plant Pathogens

Hannah Brodsky[1], Rocío Calderón[2], Douglas S. Hamilton[1,3], Longlei Li[1], Andrew Miles[4], Ryan Pavlick[5], Kaitlin M. Gold[2], Sharifa G. Crandall[4], and Natalie Mahowald[1]
1 Department of Earth and Atmospheric Sciences, Cornell University, Ithaca, NY 14853, USA
2 Plant Pathology & Plant-Microbe Biology Section, Cornell University, Geneva, NY 14456, USA
3 Now at Department of Marine, Earth and Atmospheric Science, North Carolina State, Raleigh, NC 27695, USA
4 Department of Plant Pathology and Environmental Microbiology, Pennsylvania State University, University Park, PA 16802, USA
5 Jet Propulsion Laboratory, California Institute of Technology, Pasadena, CA 91109, USA

**Abstract**
　　Pathogenic fungi are a leading cause of crop disease and primarily spread through microscopic, durable spores adapted differentially for both persistence and dispersal via soil, animals, water, and/or the atmosphere. Computational Earth System Models and air pollution models have been used to simulate atmospheric spore transport for aerial-dispersal-adapted (airborne) rust diseases, but the importance of atmospheric spore transport for soil-dispersal-adapted (soilborne) diseases remains unknown. While a few existing simulation studies have focused on *intra*continental dispersion, transoceanic and *inter*continental atmospheric transport of soilborne spores entrained in agricultural dust aerosols is understudied and may contribute to disease spread. This study adapts the Community Atmosphere Model, the atmospheric component of the Community Earth System Model, to simulate the global transport of the plant pathogenic soilborne fungus *Fusarium oxysporum* (*F. oxy*). Our sensitivity study assesses the model's accuracy in long-distance aerosol transport and the impact of deposition rate on long-distance spore transport in Summer 2020 during a major dust transport event from Northern Sub-Saharan Africa to the Caribbean and southeastern U.S. We find that decreasing wet and dry deposition rates by an order of magnitude improves representation of long-distance, trans-Atlantic dust transport. Simulations also suggest that a small number of viable spores can survive trans-Atlantic transport to be deposited in agricultural zones. This number is dependent on source spore parameterization, which we improved through a literature search to yield a global map of *F. oxy* spore distribution in source agricultural soils. Using this map and aerosol transport modeling, we show how viable spore numbers in the atmosphere decrease with distance traveled and offer a novel danger index for viable spore deposition in agricultural zones. Our work finds that intercontinental transport of viable spores to cropland is greatest between Eurasia, North Africa, and Sub-Saharan Africa, suggesting that future observational studies should concentrate on these regions.

1. **Introduction**
　　Every year, 10-15% of the world's crops are lost to disease, resulting in approximately $220 billion in losses and damages (Chakraborty & Newton, 2011; Oerke, 2006; Strange & Scott, 2005). Roughly 80% of plant diseases are caused by fungi (Tian et al., 2020). Fungi predominantly disperse through spores: small (1-60 µm), reproductive units that ensure the

organism's survival by enduring often severe environmental conditions (e.g., harsh weather conditions, lack of moisture) and persisting for long periods of time in soil (Arie, 2019; Crandall et al., 2020, 2022). Spores are capable of such extraordinary persistence that Polar science teams have discovered viable spores in Antarctic permafrost formed thousands of years ago (Kochkina et al., 2012). Plant diseases occur when a virulent fungus, a susceptible plant host, and a conducive environment converge. Understanding how fungal diseases spread and identifying agricultural zones where viable spores deposit, is vitally important to ensure global food security (Tian et al., 2020; Chakraborty & Newton, 2011; Oerke, 2006; Strange & Scott, 2005) and safeguard biodiversity by reducing fungicide usage (Brühl & Zaller, 2019) and demand for agricultural land (Williams et al., 2020).

*Fusarium oxysporum* (*F. oxy*) is a fungal disease complex and *F. oxysporum* (renamed *F. odoratissimum*) f. sp. *cubense* is the causal agent of Fusarium wilt (FW). Found on all six crop-producing continents, *F. oxy* f. sp. *cubense* is capable of infecting 100+ crops and innumerable non-crop plants. FW causes yield losses ranging from 10-60% depending on crop type, location, and infection timing (Mohd Din et al., 2020; Stoilova & Chavdarov P., 2006). For example, the emergence and subsequent spread of the highly virulent *F. oxy* f. sp. *cubense* (new designation) strain "TR-4" that causes Fusarium wilt of banana – Panama disease – has caused losses to date exceeding $500 million (Scheerer et al., 2018; Staver et al., 2020; Warman & Aitken, 2018) and may have infected ~100,000 ha of banana plantations worldwide (Ordonez et al., 2015; Ploetz, 2006, 2015). As a soil-dispersal-adapted ("soilborne") fungus, *F. oxy* spores can persist in the soil and infect plants via their roots. While soilborne fungal spores have not directly evolved for atmospheric transport, it has been well established that they can spread aerially through association with dust aerosols originating from agricultural soils (Barberán et al., 2015; Griffin et al., 2001; Kellogg et al., 2004; Rodríguez-Arias et al., 2023). Because *F. oxy* spores can remain viable in soil for at least 20 years, infected fields worldwide build up spore stocks until environmental conditions, such as high winds, are favorable for their entrainment into the atmosphere (Stover, 1962; Stover & Waite, 1960). Aerial transport occurs on a wide range of scales, from short-distance within-field transport to mid-distance regional transport to long-distance trans-continental transport (Barberán et al., 2015; Griffin et al., 2001; Kellogg et al., 2004).

Observational analyses have previously documented short-range *F. oxy* spore transport in atmospheric dust plumes (Giongo et al., 2013; Palmero et al., 2011). However, *F. oxy* capacity for mid- and long-distance transport has yet to be explored. Long-distance dispersal is key for spreading fungal diseases to new regions globally. Generally, long-range dispersal of *Fusarium* species is attributed to human and natural factors in three main ways: aerial dispersal, infested soil transport, and infected plant material movement. *F. oxy* spore size and density, 2.5-36 µm (Stover, 1962) and 1 g/cm$^3$ (Elbert et al., 2007), respectively, suggest the fungus is capable of long-distance travel using other aerosol analogies (Mahowald et al., 2014; Stover, 1962). Additionally, an *F. oxy* relative, *F. graminearum*, is well known for the intercontinental scale dispersal capacity of its aerial-dispersal-adapted ("airborne") spores (Schmale III et al., 2006, Gibberella zeae, sexual reproductive stage). Given spore morphology similarities across *Fusarium* species, it is reasonable to hypothesize that *F. oxy* may be capable of long-distance ariel transport as well (Summerell et al., 2010).

While mid- and long-distance spore transport models do exist (Prank et al., 2019; Allen-Sader et al., 2019; Sikora et al., 2014), these models are tailored for airborne spores that have different dispersal characteristics and parameters than soilborne spores. This study's goal is to use an Earth System Model for the first time to assess the possibility and patterns of soilborne fungal spore long-range transport. To accomplish this, we 1) adapted a global atmospheric model with dust aerosol tracers to include agricultural dust and spores, 2) assessed how different aerosol deposition characteristics change simulation results in comparison to observations of a large, well-observed dust event in 2020, 3) evaluated two underlying spore distribution scenarios (uniform vs. literature-informed) to assess long-range transport potential, and 4) identified global spore transport pathways.

## 2. Methods
### 2.1 Model Description.

We used the Community Atmosphere Model Version 6 (CAM6), the atmospheric component of Community Earth System Model version 2 (CESM2) (Danabasoglu et al., 2020) with a 4-mode Modal Aerosol Module (MAM4) to simulate the emission and transport of *F. oxy* spores and dust aerosols. The model simulates dust emission using a physically based dust emission scheme (Kok, Albani, et al., 2014; Kok, Mahowald, et al., 2014; Li et al., 2022) embedded in the Dust Entrainment And Deposition (DEAD) module (Zender et al., 2003). This model version includes temporal and spatial variability in dust emissions from the soil, which are typically high in unvegetated, dry areas with strong winds (Kok, Albani, et al., 2014; Kok, Mahowald, et al., 2014; Zender et al., 2003).

The model's 4 modes are Aitken, Accumulation, Coarse, and a primary carbon mode (Liu et al., 2016) with dust aerosols advected in the first three modes. Additionally, the model setup used here simulates dust aerosol in each mode as component minerals and iron using Mechanism of Intermediate complexity for Modelling Iron (MIMI) (Hamilton et al., 2019). A new dry deposition scheme (Petroff & Zhang, 2010) was recently introduced to CAM6 (Li et al., 2022) to correct the bias in simulating aerosol dry deposition velocity when using the scheme contained in the current officially released model version (Zhang et al., 2001). The dust cycle we modeled used this updated CAM6 which has been previously compared to available observations (Li et al., 2022).

### 2.2 Model Modifications

To simulate spore emissions from different geographic regions and atmospheric transport of agricultural dust and spores, we tracked all natural (non-anthropogenic) dust with one bulk tracer and independently tracked agricultural dust and spores emitted from seven distinct geographic regions (Figure 1). Understanding the relative strength of natural and anthropogenic dust sources is still an open area of research (e.g., Mahowald et al., 2010; Ginoux et al., 2012; Webb and Pierre, 2018). For this study, we assume the agricultural dust source is proportional to the crop area in each grid box based on the Climate Model Intercomparison Project crop maps for 2005 (Hurtt et al., 2020), using the same dust source mechanism described in section 3.3 (shown in Figure S1). To best match available observations, the agricultural dust source is tuned by region to have the same ratio of anthropogenic dust versus natural dust derived from satellite data in Ginoux and colleagues (2012) for each of the regions in Table S1, except for Australia. In Australia, we assume only 15% of dust is anthropogenic, consistent with other studies (e.g. Bullard et al., 2008; N. M. Mahowald et al., 2009; Webb & Pierre, 2018). The discrepancy in Australia between Ginoux et al.'s (2012) results and other studies may be caused by the large drought during the period studied in Ginoux et al., 2012.

The number of spores in any agricultural soil is not well known and so we tested two scenarios for global spore distribution in this study. In the first, a map is applied to estimate the *F. oxy* spore concentration in agricultural dust (Figure 1). This map was built using a global *F. oxy* distribution web map as a foundation (Calderón et al., 2022). We calculated an *F. oxy* survey effort ratio, which reduces the geographical biases that this web map inevitably shows. This ratio was calculated by first normalizing the number of *F. oxy* occurrence reports retrieved from the web map at the sub-country level with the total number of publications reporting plant pathogen occurrences in that region. These values were then extrapolated with the inverse distance weighting function to parts of the globe without *F. oxy* reports. Finally, the normalized occurrence rates were multiplied by the maximum concentration of *F. oxy* spores in soil based on literature (Table S2). This is the first map of its kind and thus possesses some uncertainties. It shows some spurious results from statistics of small numbers, for example, the maximum in Central Asia. In the second scenario, agricultural dust worldwide contains the maximum concentration of *F. oxy* spores. This implies that all cropland globally is highly infected with *F. oxy*. In both scenarios, *F. oxy* spores are assumed to have fixed properties (Table S2).

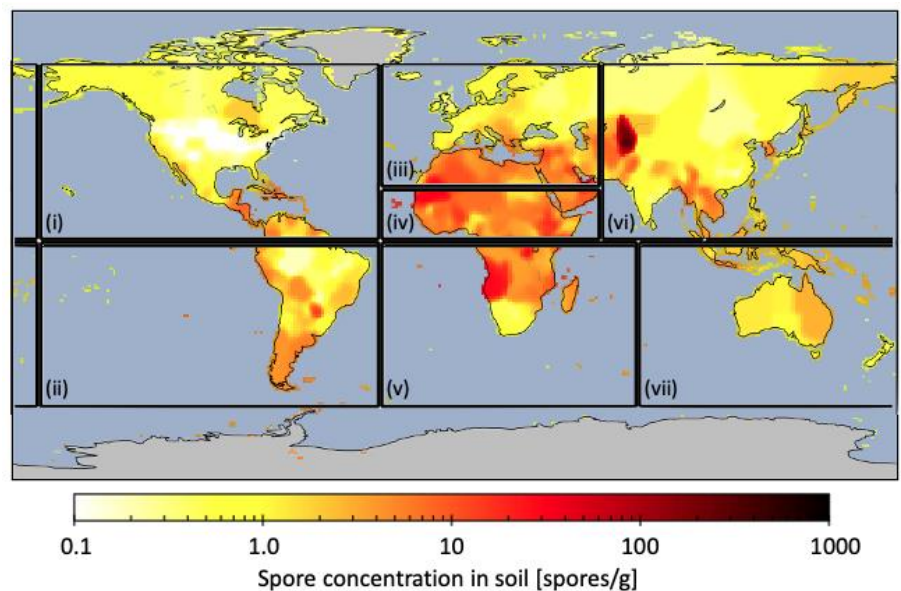

*Figure 1: Variable spore map where colors depict F. oxy spore concentration in soil [spores/g]. The map was built using the web map of global F. oxy distribution as a foundation (Calderón et al., 2022). We calculated an F. oxy survey effort ratio which reduces the geographical biases that the web map inevitably shows. This ratio was calculated by normalizing the number of F. oxy occurrence reports retrieved from the web map at the sub-country level with the total number of publications reporting plant pathogen occurrences in that region. These values were extrapolated with the inverse distance weighting function to parts of the globe without F. oxy reports. Finally, the distribution was multiplied by what we determined to be an average spore concentration for a high infestation of spores in soil: 5000 spores/g soil (Table S2). This map is applied to the production_spore_map case. Note the logarithmic scale. Geographical regions depicted are utilized throughout the study and will be referred to as: (i) North America, (ii) South America, (iii) Europe, North Africa, and Western Asia, (iv) Northern Sub-Saharan Africa, (v) Southern Sub-Saharan Africa, (vi) Central and East Asia, and (vii) Australia. All other parts of the globe are referred to as the "polar regions".*

To simulate spore deactivation during atmospheric transport, our modeling effort drew upon previous work (Meyer, Burgin, et al., 2017; Meyer, Cox, et al., 2017; Prank et al., 2019) to implement a first-order loss rate whereby spores lose their viability. Our study employed a 0.65-day e-folding lifetime, in line with empirical evidence (Maddison & Manners, 1972) and other atmospheric spore deactivation simulations (Meyer, Burgin, et al., 2017; Meyer, Cox, et al., 2017; Prank et al., 2019).

All model simulations in this study are forced with Modern-Era Retrospective Analysis for Research and Applications, Version 2 (MERRA2) reanalysis (Gelaro et al., 2017) for specific historical weather events, such as the 2020 dust event. The model is configured with a horizontal grid resolution of 0.9x1.25 degrees and 52 vertical levels.

## 2.3 Model Simulations

As a first case study, we simulated "Godzilla", a major dust storm in Summer 2020 that brought approximately 24 million tons of dust from North Africa (Emission Region) across the Atlantic Ocean towards the Caribbean and southeastern United States (Deposition Region) (see Emission Region and Deposition Region in Figure 3) (Cornwall, 2020; Masters, 2020). To capture the event and allow for spin-up, we conducted the model simulation for 14 months, starting in June 2019, and analyzed data for June 2020.

Other studies have found evidence suggesting that the deposition rate is too high in CAM6 (Meng et al., 2022). To test the sensitivity of transport to deposition rate in our model, we conducted four sensitivity studies covering June 2020: one with the default deposition rates and three with wet and/or dry deposition rates reduced by approximately one order of magnitude to increase the quantity of dust reaching the Deposition Region (Table S3). Meng and colleagues found that an order of magnitude reduction in dry deposition improved the correspondence with observations. We assume that a similar uncertainty exists for wet deposition.

To assess the impact of spore distribution on long-distance spore transport patterns, we conducted simulations with the case that best simulated the 2020 dust plume in the sensitivity study with and without the spore distribution map (simulation names production_spore_map and production_no_map, respectively, Table S3). We also conducted a sensitivity study without the spore distribution map for 40 years from 1980 through 2020 (production_40_year) to assess the implications for 2020 results and to tune production_spore_map and production_no_map based on a 0.03 global mean dust aerosol optical depth deduced in Ridley et al. (2016).

To assess global long-distance spore transport patterns, this study defines long-distance spore transport as a spore emitting from and depositing in different regions illustrated in Figure 1. We analyze the impacts of this long-distance transport by creating relative danger maps in which we multiply the deposition of long-distance transported viable spores in a subregion by the subregion's cropland area (Potapov et al., 2021) (section 3.1).

## 3. Results & Discussion
### 3.1 Sensitivity Studies for June 2020.

Similar to other studies that use the default deposition parameters (Meng et al., 2022), the modeled dust plume in this study does not penetrate as far into the Caribbean as seen in the satellite pictures (Figure S2). Of our four case scenarios, the case that reduced both the dry and wet deposition rates (dry_wet_dep_reduced) best matched satellite data of the natural dust from Northern Sub-Saharan Africa reaching the Americas (Figure 2). According to satellite data, on June 27[th] the AOD of natural dust from the Emission Region at the northeastern edge of Florida was between 0.7 and 1.6. The only case in which the natural dust AOD over Florida on June 27[th] reaches this range is dry_wet_dep_reduced. Notably, this study is primarily concerned with understanding long-distance transport and thus does not prioritize capturing mid- and short-distance transport: for example, the dry_wet_dep_reduced case appears to overestimate the quantity of dust over the Atlantic Ocean on June 23rd and 27th (Figure 2; see supplement for more discussion of the sensitivity studies).

Results from the case where variable spore distributions in the soils are applied (Figure 1, case name: production_spore_map) showed that in June 2020, a maximum of one hundred out of every one billion spores from the Emission Region that reached the Deposition Region were still viable (Figure 3). Because of the large number of spores emitted from North Africa during this extreme dust event (14 trillion spores), 13,000 viable spores from the Emission Region deposited in the Deposition Region in June 2020. Therefore, although most spores from North Africa are deactivated long before they arrive in the Americas, this initial model study *suggests that viable spores from North Africa can deposit in the Americas during extreme events*. Interestingly, when a uniform spore distribution is employed (production_no_map), a higher percent of spores in the atmosphere are viable (Figure S3), showing how the fraction of viable spores transported long-distance is sensitive to the details of the source spore spatial distribution. However, the resulting potential danger regions are similar between the runs with different spatial distributions (production_spore_map and production_no_map; Figure S4).

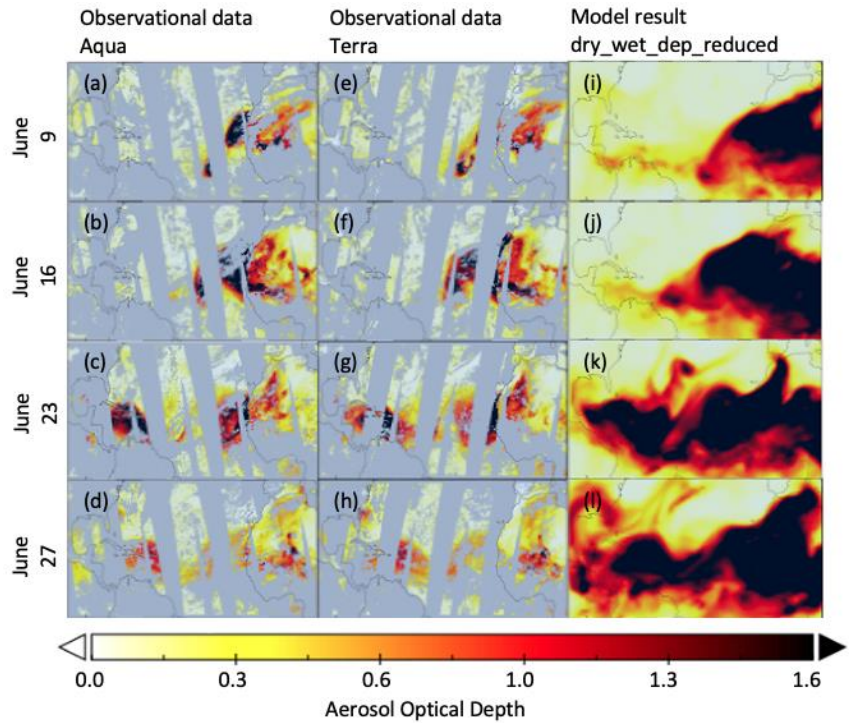

*Figure 2: Observed and modeled natural dust Aerosol Optical Depth (AOD) over the northern Atlantic Ocean throughout the transport of the June 2020 Sahara dust cloud. The colors represent different aerosol optical depth (AOD) values. The columns represent different data sources, and the rows capture the event on different days. The left and middle columns employ observational data collected from the Aqua and Terra satellites, respectively, using the Moderate Resolution Imaging Spectroradiometer (MODIS) (Levy et al., 2017). The right column uses model results from the dry_wet_dep_reduced model run (Table S3).*

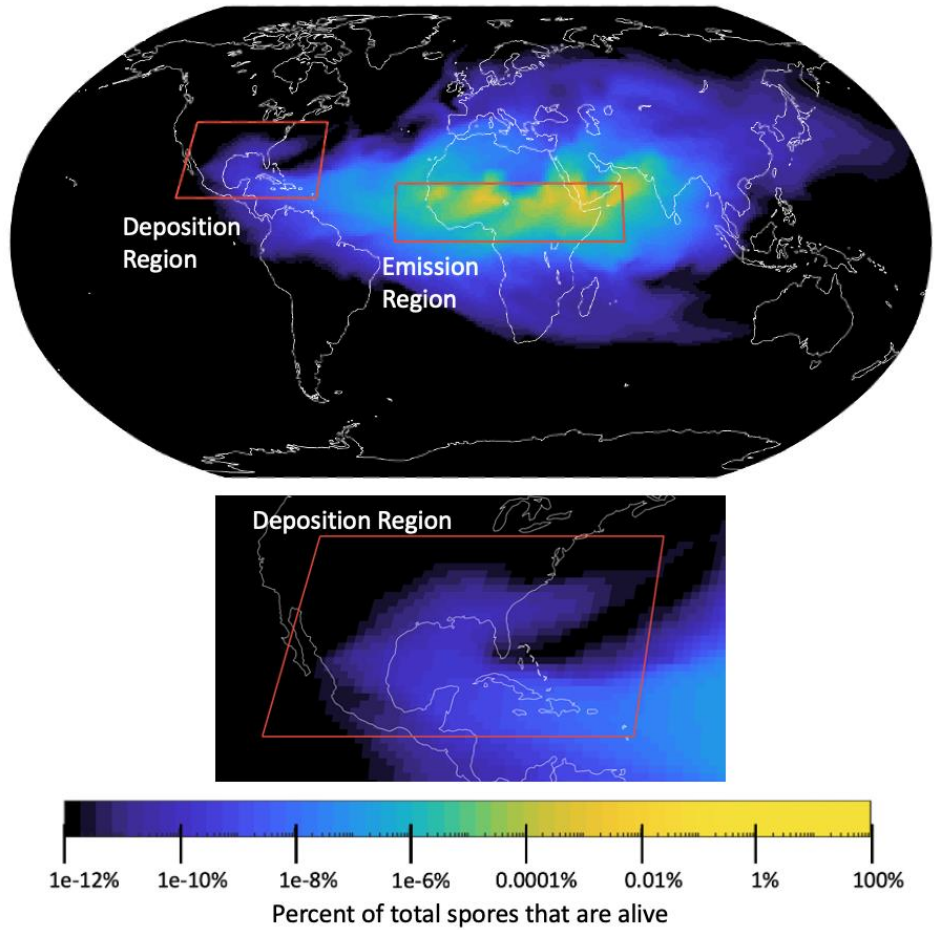

*Figure 3: The colors portray the average percent of total (viable and nonviable) spores in the atmosphere that are viable in June 2020 based on the production_spore_map case. This case employs the variable spore map and a 0.65-day e-folding lifetime. The red boxes outline the Emission Region (Northern Sub-Saharan Africa) and Deposition Region (southeastern North America and the Caribbean) used when examining long-distance transport during the large June 2020 Sahara dust cloud. Note the logarithmic scale.*

Next, we explore the implications of viable spore deposition by examining the overlap in viable deposition with cropland locations in the Americas. To do this, we introduce a danger index for such regions. The danger index is defined in this study as the product of the number of live spores deposited per million hectares in June 2020 and the percent of land used for crops in each grid cell. Because different *F. oxy* species can impact many crops, we include all croplands in this calculation (Figure 4a). Since the plume reaches Central America and the Caribbean Islands before the continental U.S. (Figure 2) it deposits more spores onto the former (Figure 4a). We find areas of particularly high danger of live spore deposition in cropland include Mexico, Cuba, Haiti, and the Dominican Republic.

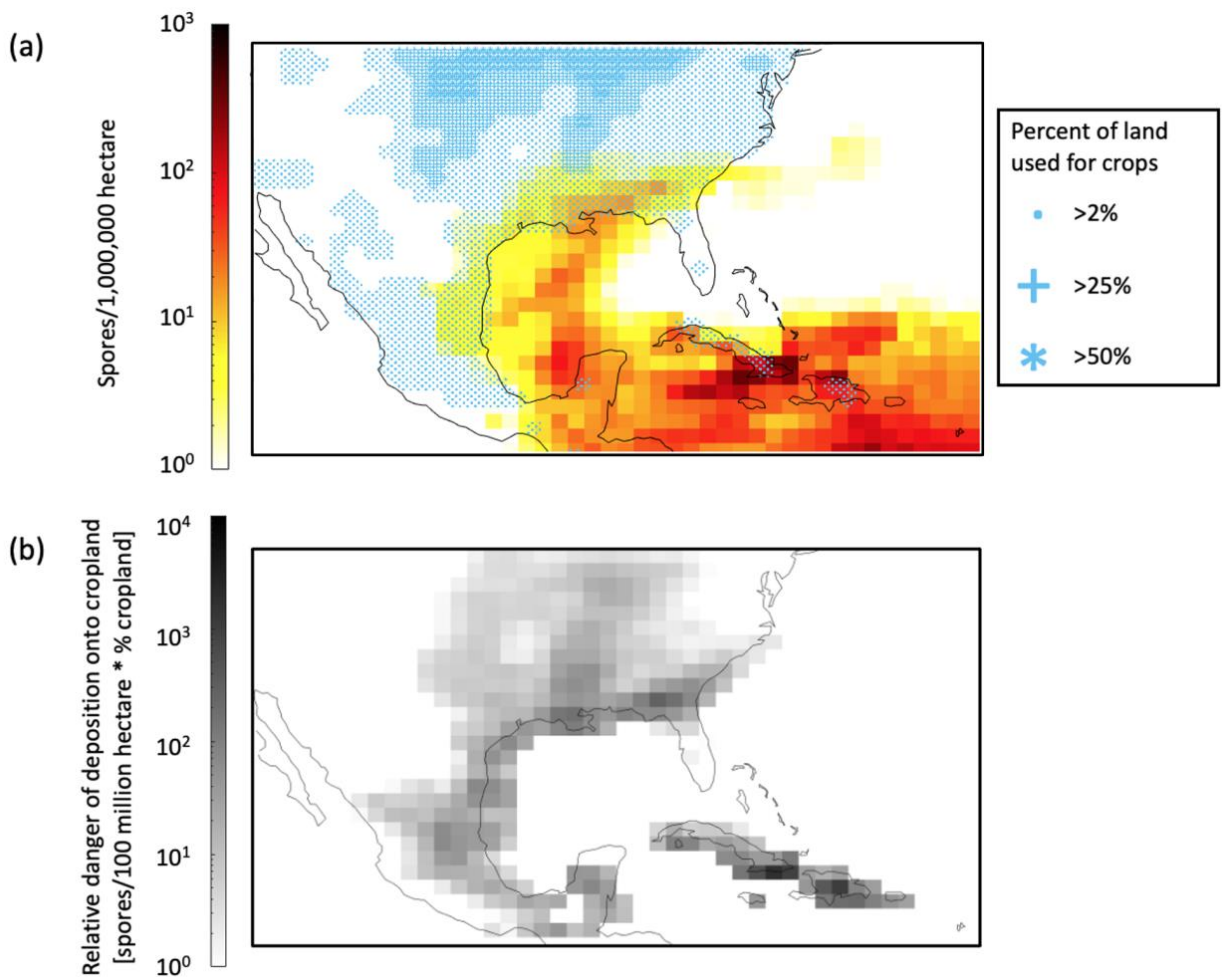

*Figure 4: (a) Yellow-red colors represent the number of live spores deposited in the Deposition Region per million hectares in June 2020 as modeled by production_spore_map which includes a variable spore map. Plot only considers spores from the Emission Region. Stippling shows the percent of land used for crops based on Potapov et al. (2021). (b) Shading represents the danger of viable long-distance transported spore deposition, defined in this study as the multiplication of the number of live spores deposited per million hectares in June 2020 and the percent of land used for crops in each grid cell. The number of live spores is model output from production_spore_map run and only considers spores from the Emission Region. Note the logarithmic scale for (a) and (b).*

### 3.2 Global Implications

Results from the model (using case production_spore_map) show that on average in 2020, similar to just June 2020 (Figure 3), a larger percentage of total spores deposited were long-distance transported (58%) than viable spores deposited (2%; long-distance transport is defined as transport between two different regions in Figure 1). There is substantial heterogeneity between regions' estimated spore emissions as well as variation in spore deposition location (Figure 5). The number of transported viable spores is much lower than that of total spores (Figure 5b). Long-distance viable spore transport is dominated by interactions between western Eurasia, North Africa, and Northern Sub-Saharan Africa (Regions III, IV, and VI); 96% of viable long-range transported spores either originate or end up in one of these regions (Figure 5). The connection between these regions is in part due to proximity (Figure 1). Proximity also directly increases long-range viable spore transport; higher percentages of spores are viable closer to their emission location, so viable spore transport is more likely to occur over shorter distances. The large amount of modeled spore emissions from the regions (74% of global spore emissions) also contributes to the prevalence of all spore transport, including long-distance viable spore transport, in the regions. Relatively large spore emissions are in part caused by high spore concentrations in the crop soils in western Eurasian and African regions (Figure 1).

These regions' aerial interconnectedness may lead to shared genetic material across the regions. Additionally, when combined with the cropland prevalence and high spore emissions in the region, the connectivity produces a high rating for the danger of viable long-distance transport deposition on cropland across Eurasia-Africa regions (Figure 6). Notably, this danger of spore deposition does not imply infection risk, which requires consideration of spore pathogenicity, crop susceptibility, and conducive environmental conditions such as local site microclimate and topography as well as changes in temperature (Delgado-Baquerizo et al., 2020).

Sub-Saharan Africa (Regions IV and V) appears to be a hub for long-distance viable spore transport (Figures 5 and 6). In this study, relative long-distance viable spore transport frequency is measured as a percent of total long-distance viable spore transport between all regions of the world (Figure 6). The model estimates that Sub-Saharan Africa is connected to five other crop-producing continents via long-distance viable spore transport above the 0.0001% (one in a million) frequency threshold: more than other regions. Based on the simulation, we estimate that on average 53% of all viable spore deposition and 14% of long-distance transported viable spore deposition originates in Sub-Saharan Africa. Additionally, the model estimates that in 2020, over 7 billion spores emitted from Sub-Saharan Africa were long-distance transported per month; Sub-Saharan Africa's high spore deposition and emission rates could be important for disease spread.

In addition to transport within and between the Eurasian and African continents, the model suggests other pathways exist for long-distance viable spore transport (Figure 5). Viable spores could be transported long-distance from South America to North America and from Australia to South America. Viable spores emitted from Asia could also deposit in North America. Other pathways for long-distance viable spore transport (Figure 5b) each represent less than 0.0001% of global viable spore emissions.

Including a spore distribution map (instead of assuming uniform maximum spore infection rates) in the model significantly reduces the number of spores emitted and thus long-distance transport potential, but the patterns of long-distance viable spore transport remain similar (seen by comparing production_spore_map and production_no_map; Figure 6 versus S5).

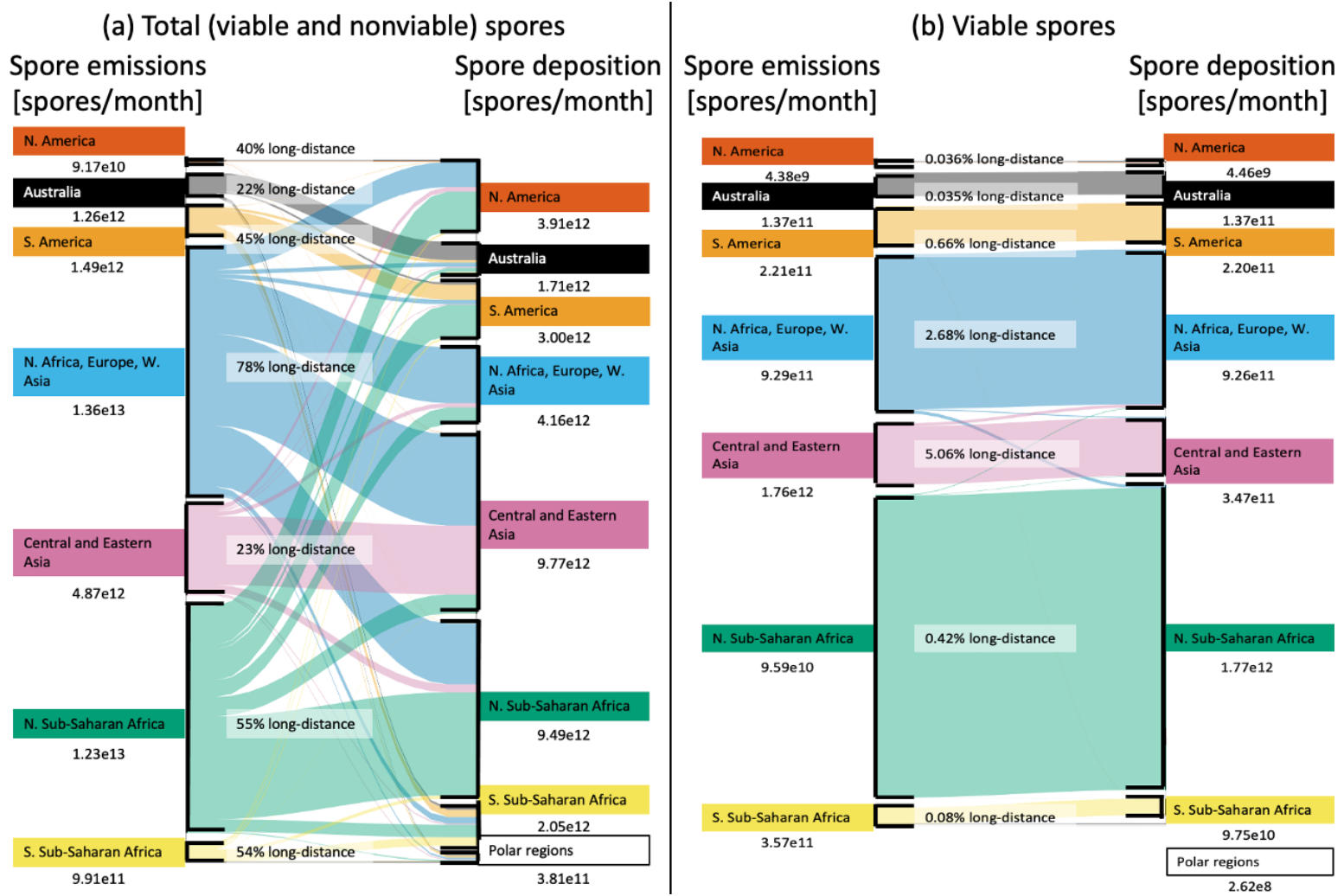

Figure 5: The Sankey diagrams depict the global atmospheric spore flow between geographical regions as computed in the production_spore_map model, which includes a variable spore map. While the polar regions do not emit spores, some spores deposit in those areas. Transport is measured in units of average monthly spores transported in 2020. Also included is the percentage of spores from each emission region that are long-distance transported, defined as being emitted from and deposited in different regions. (a) depicts total (viable and nonviable) spores while (b) depicts viable spores. In (b), only spores that deposit before deactivating are considered in the calculations for spore emission, spore deposition, and percent long-distance transported. The Sankey diagrams were made by adapting code by Carmeli (2022).

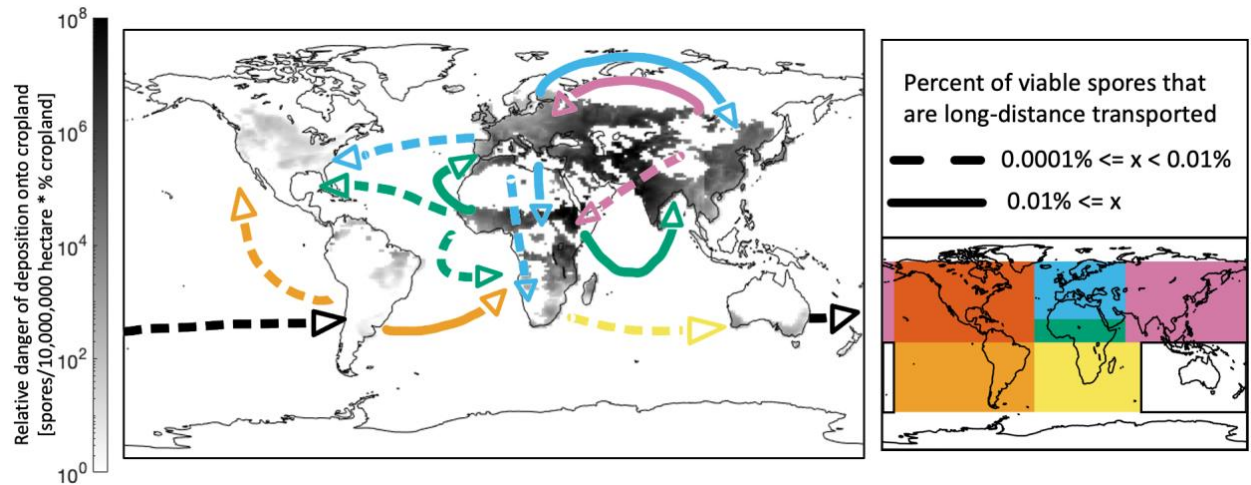

*Figure 6: Gray shading depicts the danger of viable long-distance transported spore deposition onto cropland, defined in this study as the multiplication of the number of live, long-distance transported spores deposited in the Deposition Region per 10 million hectares and the percent of land used for crops in each grid cell. The number of live spores is model output from the production_spore_map run which includes a variable spore distribution map. Percent of land used for crops is taken from Potapov et al. (2021). Arrows depict the most utilized pathways in 2020 for long-distance viable spore transport, defined as spores emitted from and depositing in different regions while remaining viable. Spore transport is measured as a percent of total long-distance viable spore transport between all regions of the world. For example, based on production_spore_map, in 2020 an average of 3.5e12 viable spores long-distance transported per month between all regions of the world, and of those spores, 1.7e8 traveled from Central and East Asia to Northern Sub-Saharan Africa. Therefore, the Central and East Asia - Northern Sub-Saharan Africa pathway percentage is 0.004906%. Only percentages reaching the 0.0001% threshold are depicted as arrows. For instance, there are spores emitted from North America that undergo long-distance viable spore transport, but no long-distance transport pathways originating in North America exceed the threshold, so no arrows originate in the region. All calculations only consider spores that deposit before deactivating. Arrow colors correspond to emission region, and arrow line types correspond to transport percentage. If the transport percentage is greater than 0.01%, solid lines are used for arrows; otherwise, dashed lines are used for arrows.*

## 4. Conclusions

This work suggests that while regional atmospheric transport of viable soilborne spores is far more common, long-distance transport of viable soilborne spores may also be possible. We find that spores are most likely to be transported between Eurasia and Africa, and thus future observational studies focused on spore transport in dust plumes from these regions are most likely to help quantify soilborne spore transport.

Our modeling study represents the first step towards quantitively understanding long-distance soilborne spore transport, especially during large meteorological phenomena such as intra/intercontinental dust storms. Understanding long-distance plant pathogen transport could benefit from future research by quantifying: the spore fraction in agricultural dust, the agricultural land spatial distribution, particle atmospheric residence time (Textor et al., 2006; Meyer, Burgin, et al., 2017; Meyer, Cox, et al., 2017; Prank et al., 2019), particle deposition rate, particle size (Mahowald et al., 2014; Tegen & Lacis, 1996; Zender et al., 2003), shape (Huang et al., 2020), and composition; and spore deactivation rate (Maddison & Manners, 1972, 1973; Meyer, Burgin, et al., 2017; Meyer, Cox, et al., 2017; Prank et al., 2019; Visser et al., 2019). Model parameterizations of aerosol transport processes and environmental factors alter deposition patterns (Hamilton et al., 2019; Kok, 2011; Kok et al., 2021; Tegen & Lacis, 1996; Zender et al., 2003) and thus such model refinements would improve future spore deposition risk predictive capabilities.

This study lays the groundwork for modeling long-distance viable spore transport and deposition, with the aim of building an operational, real-time global surveillance system of long-distance plant pathogen transport risk (Carvajal-Yepes et al., 2019). Our transport model could be parameterized for modeling the dispersal of other types of plant pathogens where spore biology (e.g., morphology, viability) is already known. Specific to *F. oxy*, additional relevant factors for understanding plant infection include the pathogenicity of spores transported, susceptibility of crops to deposited *F. oxy* strains, and environmental conditions during deposition. This study models *F. oxy* transport, but not all *F. oxy* strains are pathogenic, with some living saprotrophic or endophytic ecological lifestyles. While *F. oxy* species have a wide plant host range, each pathogenic *F. oxy* strain has a narrow pathogenic specificity to host species, leading to the *formae speciales* concept to group strains by hosts that exhibit disease symptoms. Additionally, some *formae speciales* are further divided into races because of cultivar-level specialization. Thus, infection by long-distance transport depends on whether crops in the spore deposition region are susceptible to the pathogens transported from the emission region. Such alignment may be more common when transport occurs between regions with similar environmental conditions. Once pathogenic spores land on susceptible crops, environmental conditions also must be conducive to infection and disease development (Sussman & Douthit, 1982). Others have shown, through quantifying the microbial taxa found in dust samples, that fungi are the dominant taxa that disperse during dust storms from Sub-Saharan Africa to central America (Waters et al., 2020). Many fungal pathogen spores possess melanized cell walls to shield them from the negative effects of ultraviolet light, which can render them non-viable (Nakpan et al., 2019). Future studies should conduct empirical viability analyses that would improve our understanding of whether pathogenic spores could survive long-distance transport. These studies could culture directly from air filter samples and would shed light on whether spores survive their dispersal journeys and retain the ability to germinate (Chen & Séguin-Swartz, 2002).

## Data Access Statement

The data that support the findings of this study are openly available at the following DOI: https://doi.org/10.7298/ddgx-ht24

## Acknowledgments

HKB, RC, KMG, SGC, RP, and NMM would like to acknowledge the support of NASA (80NSSC20K1533). Additionally, a portion of this research was carried out at the Jet Propulsion Laboratory, California Institute of Technology, under a contract with the National Aeronautics and Space Administration (80NM0018D0004). We would also like to acknowledge high-performance computing support from Cheyenne (doi:10.5065/D6RX99HX) provided by NCAR's Computational and Information Systems Laboratory, sponsored by the National Science Foundation.

**Supplementary material for "Assessing Long-Distance Atmospheric Transport of Soilborne Plant Pathogens"**

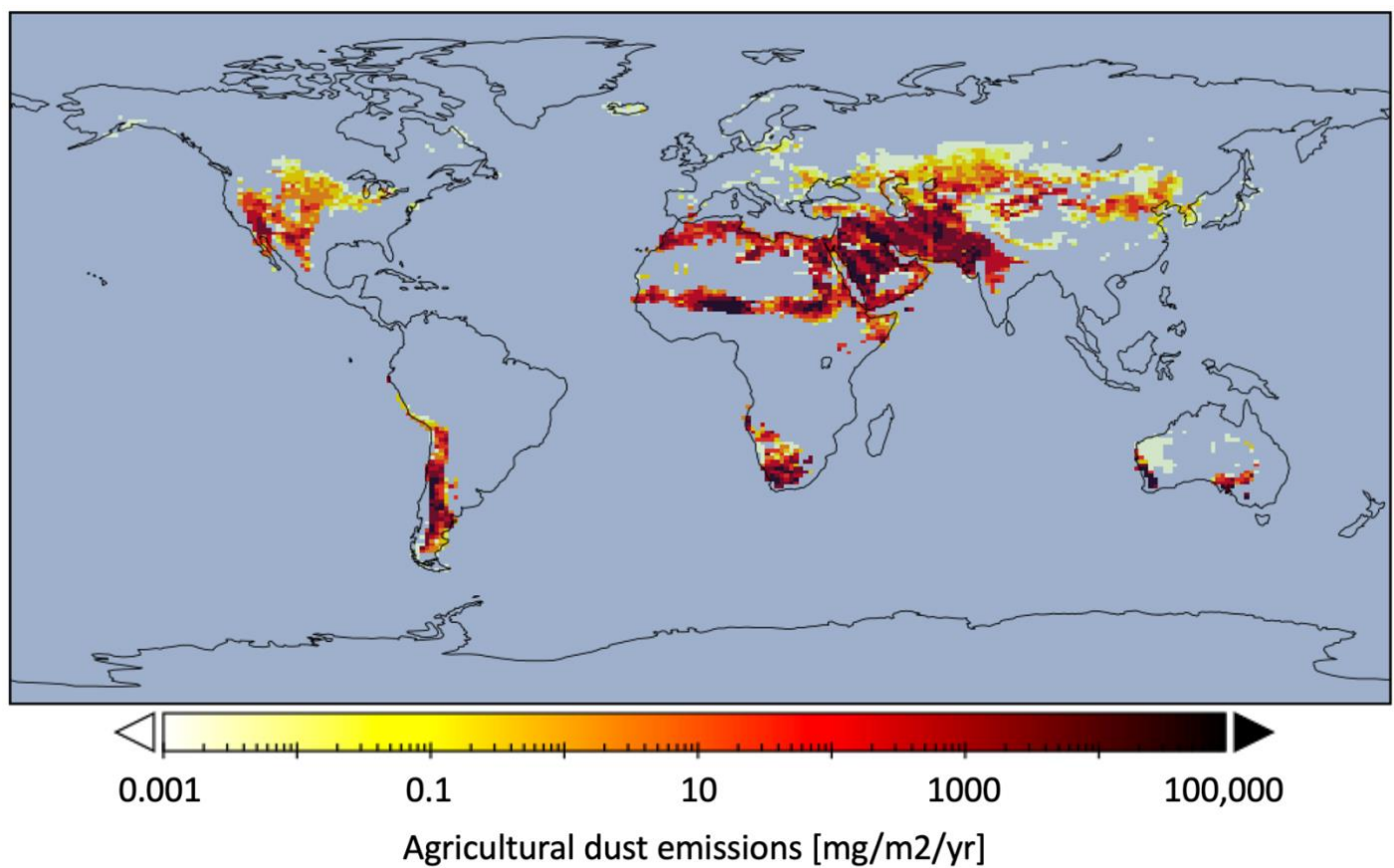

*Supplementary Figure 1: Model output for agricultural dust emissions. Colors represent tuned agricultural dust emissions globally (mg/m$^2$/yr). Input for agricultural dust source was developed by assuming the agricultural dust source is proportional to the crop area in each grid box, using the same dust source mechanism described in section 3.3. To best match available observations, the agricultural dust source is tuned by region to be proportional to the amount of anthropogenic dust derived from satellite in* Ginoux et al., 2012 *for each of the regions in Supplementary Table 3, except for Australia; in Australia, we assume only 15% of dust is anthropogenic, consistent with other studies (e.g., Bullard et al., 2008; Mahowald et al., 2009; Webb & Pierre, 2018). The discrepancy in Australia between Ginoux et al.'s (2012) results and other studies may be caused by the large drought during the time period studied in Ginoux et al., 2012.*

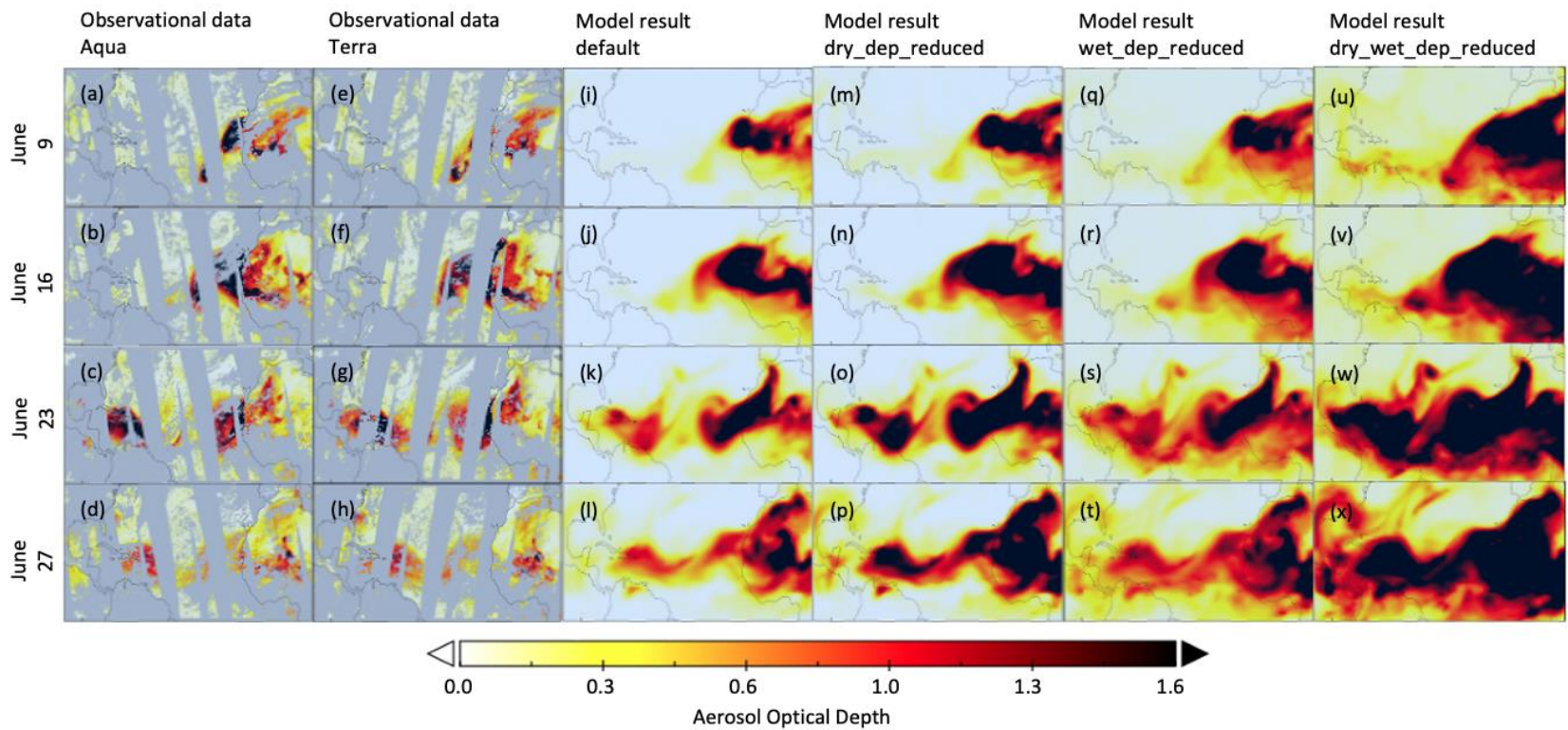

*Supplementary Figure 2: Observed and modeled natural dust Aerosol Optical Depth (AOD) over the northern Atlantic Ocean throughout the transport of the June 2020 Sahara dust cloud. Colors represent different aerosol optical depth (AOD) values. The columns represent different data sources, and the rows capture the event on different days. The first and second columns employ observational data collected from the Aqua and Terra satellites, respectively, using the Moderate Resolution Imaging Spectroradiometer (MODIS) (Levy et al., 2017). The third through sixth columns use model results from the default, dry_dep_reduced, wet_dep_reduced, and dry_wet_dep_reduced model runs, respectively (Supplementary Table 2).*

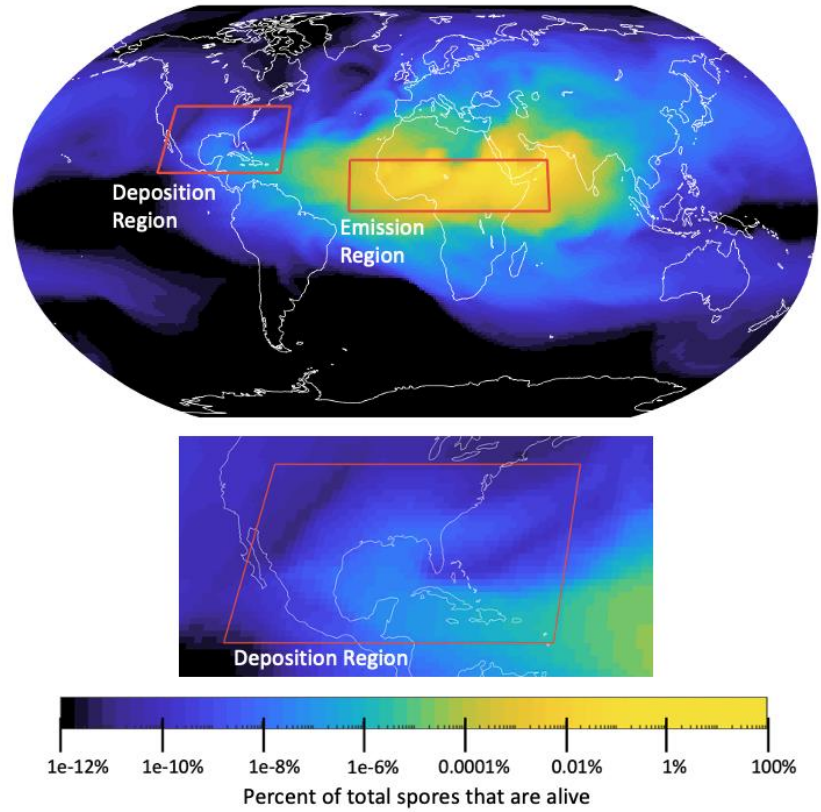

*Supplementary Figure 3: The colors portray the average percent of total (viable and nonviable) spores that are viable in June 2020 based on the production_no_map case. This case assumes a uniform spore distribution and a 0.65-day e-folding lifetime. The red boxes outline the Emission Region (Northern Sub-Saharan Africa) and Deposition Region (southeastern North America and the Caribbean) used when examining long-distance transport during the large June 2020 Sahara dust cloud. Note the logarithmic scale.*

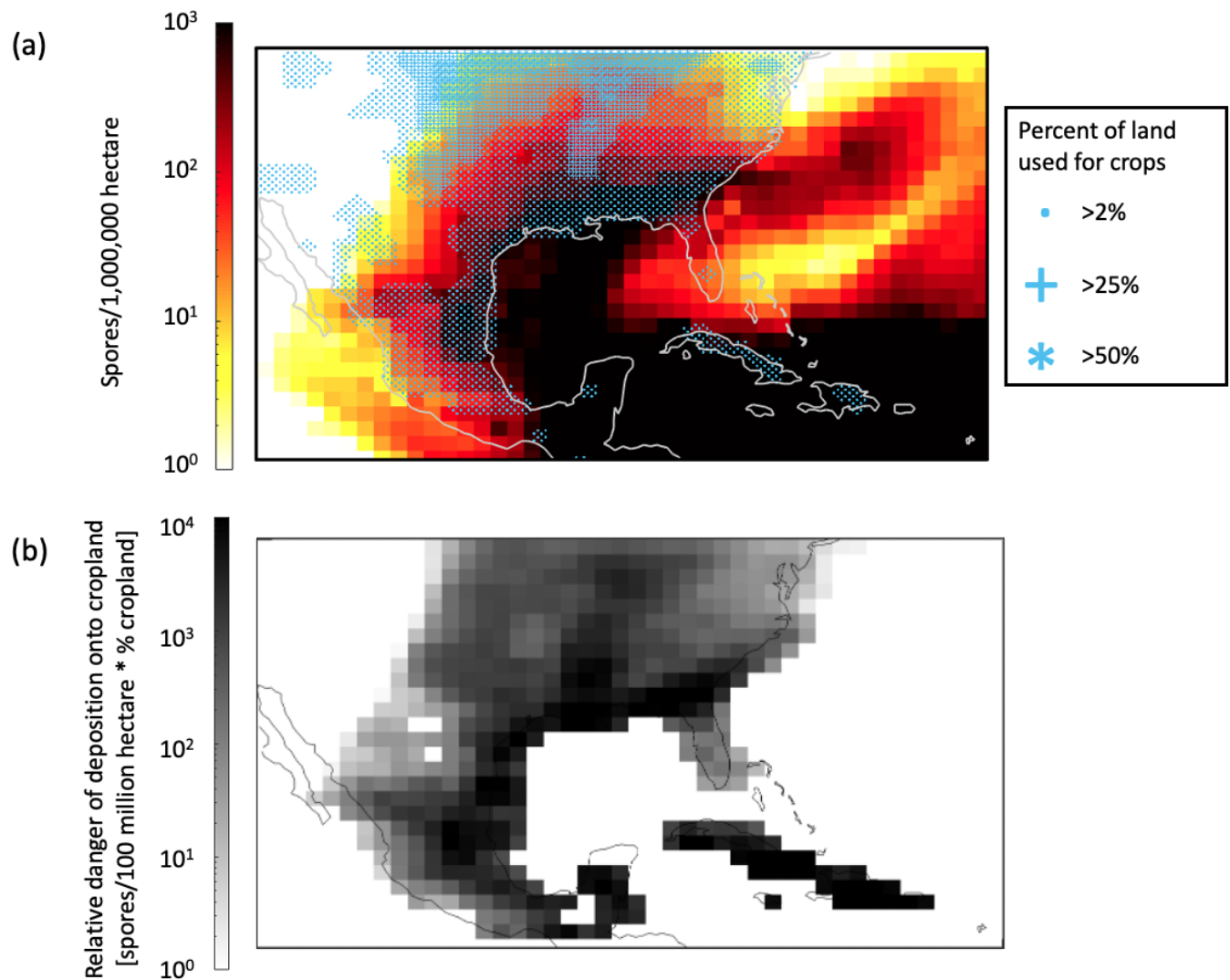

*Supplementary Figure 4: (a) Yellow-red colors represent the number of live spores deposited in the Deposition Region per million hectares in June 2020 as modeled by production_no_map which includes uniform maximum spore distribution. Plot only considers spores from the Emission Region. Stippling shows the percent of land used for crops based on Potapov et al. (2021). (b) Shading represents the danger of viable long-distance transported spore deposition, defined in this study as the multiplication of the number of live spores deposited per million hectares in June 2020 and the percent of land used for crops in each grid cell. Number of live spores is model output from production_no_map run and only considers spores from the Emission Region. Note the logarithmic scale for (a) and (b).*

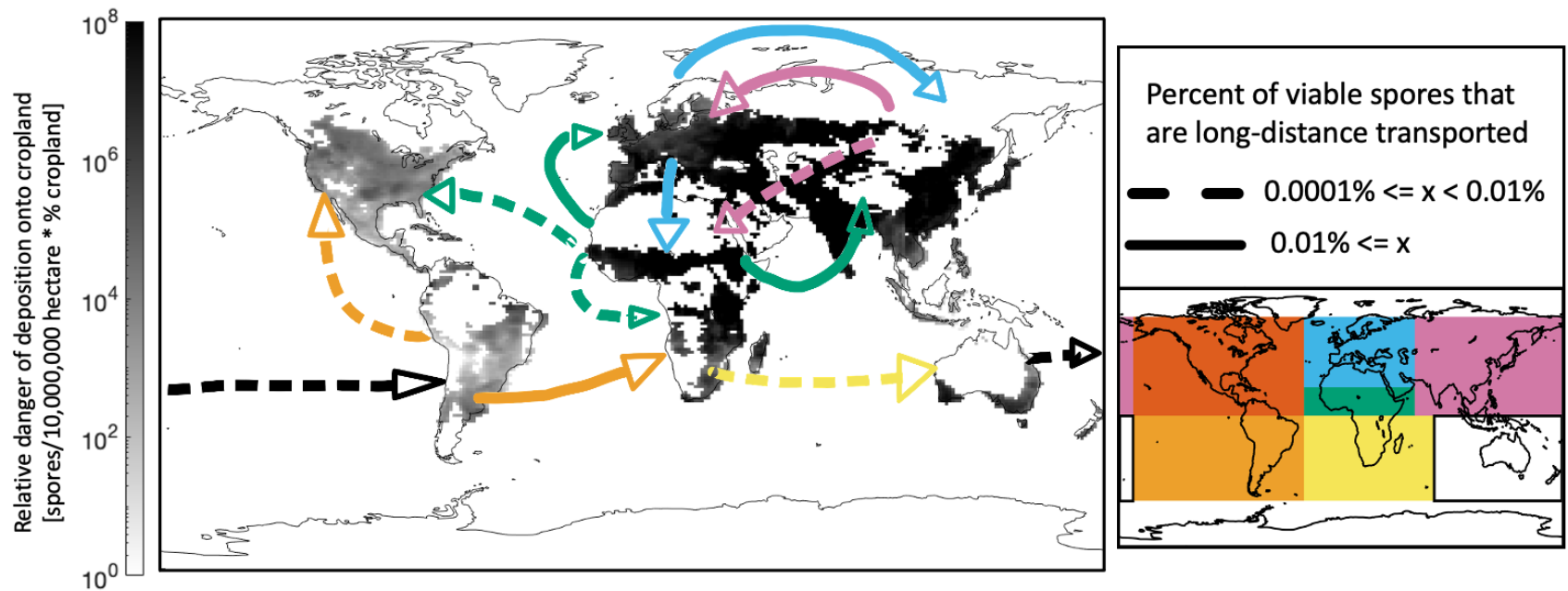

*Supplementary Figure 5: Gray shading depicts the danger of viable long-distance transported spore deposition onto cropland, defined in this study as the multiplication of the number of live, long-distance transported spores deposited in the Deposition Region per 10 million hectares and the percent of land used for crops in each grid cell. Number of live spores is the model output from production_no_map run which assumes uniform maximum spore distribution. Percent of land used for crops is taken from Potapov et al. (2021). Arrows depict the most utilized pathways in 2020 for long-distance viable spore transport, defined as spores emitted from and depositing in different regions while remaining viable. Spore transport is measured as a percent of total long-distance viable spore transport between all regions of the world. For example, based on production_no_map, in 2020 an average of 4.5e15 viable spores long-distance transported per month between all regions of the world, and of those spores, 3.0e11 traveled from Central and East Asia to Northern Sub-Saharan Africa. Therefore, the Central and East Asia - Northern Sub-Saharan Africa pathway percentage is 0.00658%. Only percentages reaching the 0.0001% threshold are depicted as arrows. For instance, there are spores emitted from North America that undergo long-distance viable spore transport, but no long-distance transport pathways originating in North America exceed the threshold, so no arrows originate in the region. All calculations only consider spores that deposit before deactivating. Arrow colors correspond to emission region, and arrow line types correspond to transport percentage. If the transport percentage is greater than 0.01%, solid lines are used for arrows; otherwise, dashed lines are used for arrows.*

**Supplementary Table 1:** Anthropogenic Percentage of Dust [a]

|  | North America | South America | North Africa | South Africa | Western Asia | Central Asia | Eastern Asia | Australia |
|---|---|---|---|---|---|---|---|---|
| % Anthropogenic – Ginoux et al., 2012 | 54 | 41 | 8.2 | 54 | 30 | 45 | 40 | 76[b] |
| % Anthropogenic - Tuned Run | 49 | 40 | 8.4 | 59 | 32 | 48 | 38 | 15 |

[a] Average emissions over the regions specified in Ginoux et al., 2012 were calculated and crop area effectiveness as a dust source was modified to create the tuned anthropogenic crop dust source.
[b] Other studies have found a lower percent agricultural dust for Australia (Bullard et al., 2008; Mahowald et al., 2009; Webb & Pierre, 2018)

**Supplementary Table 2.** *F. oxysporum* Spore Characteristics

| Characteristic | Average (value applied in model) | Minimum | Maximum | Uncertainty (orders of magnitude) | Sources |
|---|---|---|---|---|---|
| Concentration of Spores in Soil (Spores/g Soil) | 5000 | 500 | 10,000 | ± 1 | (Naiki & Morita, 1983; Smith & Snyder, 1971) |
| Density of Individual Spore (g/cm$^3$) | 1 | 1 | 1 | 0 | (Buller, 1909; Elbert et al., 2007; Trail et al., 2005) |
| Size of Individual Spore (μm$^3$) | 570[a] | 30[a] | 4511[a] | ± 1 | (Stover, 1962) |
| Weight of Individual Spore (pg) | 570[b] | 30[b] | 4511[b] | ± 1 | n/a |

[a] Size assumes sphericity. [b] Calculated using density*size.

**Supplementary Table 3:** Model simulations

| | Run name | Dates | Model features |
|---|---|---|---|
| Sensitivity studies | default | June 2019 - August 2020 | – Uniform spore distribution<br>– Default deposition variables |
| | dry_dep_reduced | June 2019 - August 2021 | – Uniform spore distribution<br>– Reduce gravitational settling and turbulent removal rate by an order of magnitude |
| | wet_dep_reduced | June 2019 - August 2022 | – Uniform spore distribution<br>– Reduce precipitation loss rate by an order of magnitude |
| | dry_wet_dep_reduced | June 2019 - August 2023 | – Uniform spore distribution<br>– Reduce gravitational settling, turbulent removal, and precipitation by an order of magnitude |
| Production runs | production_no_map | June 2019 - January 2021 | – Uniform spore distribution<br>– Reduce gravitational settling, turbulent removal, and precipitation by an order of magnitude |
| | production_spore_map | June 2019 - January 2021 | – Variable spore distribution<br>– Reduce gravitational settling, turbulent removal, and precipitation by an order of magnitude |
| | production_40_year | January 1980 - December 2021 | – Uniform spore distribution<br>– Reduce gravitational settling, turbulent removal, and precipitation by an order of magnitude |

**Comparing cases with different deposition rates**

Supplementary Figure 6 shows dust AOD at 20°N during the summer 2020 dust event. 20°N was the latitude of the maximum plume AOD on June 9, 2020, the day the plume began to cross the western coast of Africa. Thus, we assume 20°N to be the latitudinal "middle" of the plume throughout the course of the event. In this analysis, the AOD from observational data and from model results has been averaged over 5° to account for the difference in grid size between the model and observational data.

As seen in Supplementary Figure 6, all four cases produce relatively similar AOD values at 20°N. The dry_wet_dep_reduced case consistently has the highest AOD values (Supplementary Figure 6; Supplementary Table 4a, 4b) which is expected since this case has the lowest deposition rate. Some AOD extremes in the observational data are not captured in any of the model simulations, for example on June 23 and June 25 (Supplementary Figure 6).

This study also provides insight into the role of wet and dry deposition in aerosol dust transport. As seen in Supplementary Figure 2, compared to the default case, decreasing wet deposition tends to increase the spatial distribution of aerosol transport. In contrast, decreasing dry deposition primarily increases absolute spore transport distance. Decreasing both wet and dry deposition increases the distance and spatial distribution of aerosol transport. Particle residence time is negatively related to deposition rate; therefore, similar to other studies, this research suggests that particle residence time is positively related to particle travel distance and deposition velocity (Textor et al., 2006).

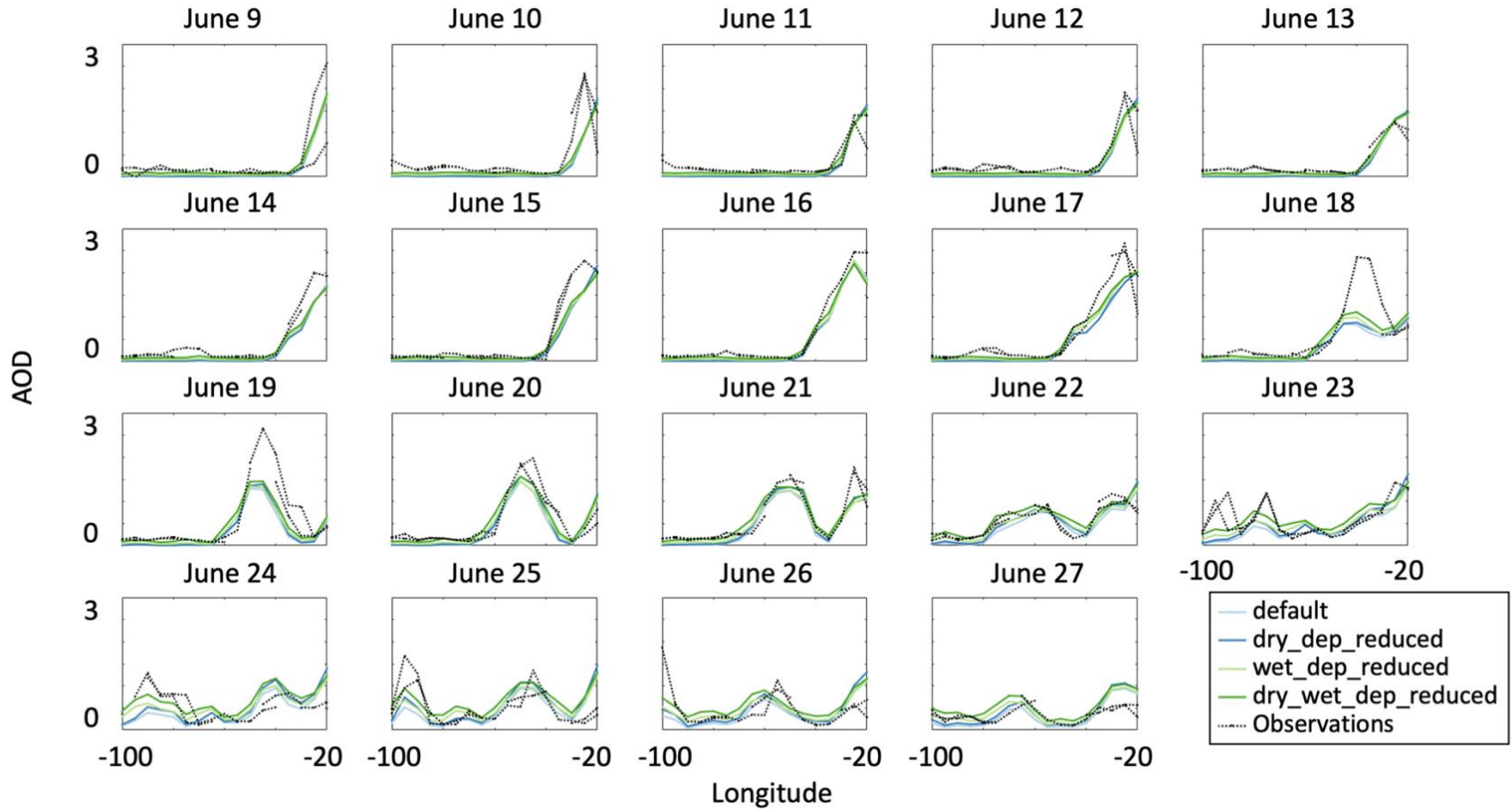

*Supplementary Figure 6: Aerosol optical depth of natural dust at 20°N between longitudes 100° W and 20°W from June 9 through June 27 of 2020- the period of a large Sahara dust event that brought dust from Northern Sub-Saharan Africa to the southeastern U.S. and Caribbean. Colored lines represent output from different simulations of the event (Supplementary Table 3). Black dashed lines employ observational data collected from the Aqua and Terra satellites using the Moderate Resolution Imaging Spectroradiometer (MODIS)* (Levy et al., 2017).

**Supplementary Table 4a: Longitude and AOD of plume peak at 20°N - June 17 - June 24**

|  | June 17 | | June 18 | | June 19 | | June 20 | |
| --- | --- | --- | --- | --- | --- | --- | --- | --- |
|  | Longitude | Value | Longitude | Value | Longitude | Value | Longitude | Value |
| default case | -20 | 2.06 | -40 | 0.82 | -45 | 1.27 | -50 | 1.45 |
| dry_dep_reduced case | -20 | 2.02 | -40 | 0.87 | -45 | 1.39 | -50 | 1.56 |
| wet_dep_reduced case | -20 | 2.06 | -40 | 0.99 | -45 | 1.32 | -50 | 1.46 |
| dry_wet_dep_reduced case | -20 | 2.01 | -40 | 1.11 | -45 | 1.45 | -50 | 1.56 |
| Aqua | -25 | 2.67 | -40 | 2.31 | - | - | -50 | 1.86 |
| Terra | -25 | 2.48 | -40 | 2.35 | -45 | 2.65 | -45 | 1.98 |

|  | June 21 | | June 22 | | June 23 | | June 24 | |
| --- | --- | --- | --- | --- | --- | --- | --- | --- |
|  | Longitude | Value | Longitude | Value | Longitude | Value | Longitude | Value |
| default case | -50 | 1.19 | -60 | 0.74 | -60 | 0.48 | -90 | 0.39 |
| dry_dep_reduced case | -50 | 1.28 | -60 | 0.81 | -80 | 0.58 | -90 | 0.52 |
| wet_dep_reduced case | -50 | 1.22 | -60 | 0.78 | -80 | 0.58 | -90 | 0.61 |
| dry_wet_dep_reduced case | -50 | 1.32 | -60 | 0.91 | -80 | 0.78 | -90 | 0.81 |
| Aqua | -50 | 1.51 | -55 | 0.93 | -90 | 1.20 | -90 | 1.18 |
| Terra | -50 | 1.60 | -60 | 0.92 | -75 | 1.19 | -90 | 1.30 |

Note: Grey boxes flag case with highest peak AOD on a given day

**Supplementary Table 4b: Longitude and AOD of plume peak at 20°N - June 25 - June 27**

|  | June 25 | | June 26 | | June 27 | |
|---|---|---|---|---|---|---|
|  | Longitude | Value | Longitude | Value | Longitude | Value |
| default case | -95 | 0.52 | -100 | 0.32 | -100 | 0.16 |
| dry_dep_reduced case | -95 | 0.74 | -100 | 0.47 | -100 | 0.24 |
| wet_dep_reduced case | -95 | 0.68 | -100 | 0.52 | -100 | 0.33 |
| dry_wet_dep_reduced case | -95 | **0.94** | -100 | **0.73** | -100 | **0.48** |
| Aqua | -95 | 1.68 | - | - | -100 | 0.36 |
| Terra | -90 | 1.13 | -100 | 1.87 | -100 | 0.45 |

Note: Grey boxes flag case with highest peak AOD on a given day